\documentclass[11pt]{article}
\usepackage{amsmath,amssymb,amsfonts}
\usepackage{latexsym}
\usepackage{pstricks}  
\def\pcirc{
\stackrel{   \raisebox{-4pt}{\scalebox{.85}{  $\circ$ }}  }
             {   \raisebox{-1pt}{\scalebox{1.05}{  $p$ }}   } }
\begin{document}
\begin{titlepage}
\begin{flushright}
UTAS-PHYS-2008-\hspace*{.2cm}\\
June 2008\\
\end{flushright}
\begin{centering}
 
{\ }\vspace{0.5cm}
 
{\Large\bf Constraint quantisation of a worldline system}

\vspace{5pt}

{\Large\bf  invariant under reciprocal relativity. II.}

\vspace{1.8cm}

P. D. Jarvis\footnote{Alexander von Humboldt Fellow} and
S. O. Morgan\footnote{Australian Postgraduate Award}

\vspace{0.3cm}

{\em School of Mathematics and Physics}\\
{\em University of Tasmania, Private Bag 37}\\
{\em Hobart, Tasmania 7001, Australia }\\
{\em E-mail: {\tt Peter.Jarvis@utas.edu.au}, {\tt Stuart.Morgan@utas.edu.au}}

\clearpage
\thispagestyle{empty}
\begin{abstract}
\noindent
We consider the world-line quantisation of a system invariant under the symmetries of reciprocal relativity. Imposition of the first class constraint, the generator of local time reparametrisations, on physical
states enforces identification of the world-line cosmological constant with a fixed value of the quadratic Casimir of the quaplectic symmetry group $Q(3,1)\cong U(3,1)\ltimes H(4)$, the
semi-direct product of the pseudo-unitary group with the Weyl-Heisenberg group. In our previous paper, J Phys A \textbf{40} (2007) 12095--12111, the `spin' degrees of freedom were handled as covariant oscillators, leading to a unique choice of cosmological constant, required for projecting out negative-norm states from the physical gauge-invariant states. In the present paper the spin degrees of freedom are treated as standard oscillators with positive norm states (wherein Lorentz boosts are not number-conserving in the auxiliary space; reciprocal transformations are of course not spin-conserving in general). As in the covariant approach, the spectrum of the square of the energy-momentum vector is continuous over the entire real line, and thus includes tachyonic (spacelike) and null branches. Adopting standard frames, the Wigner method on each branch is implemented, to decompose the auxiliary space into unitary irreducible representations of the respective little algebras and additional degeneracy algebras. The physical state space is vastly enriched as compared with the covariant approach, and contains towers of integer spin massive states, as well as unconventional massless representations, with continuous Euclidean momentum and arbitrary integer helicity. 
\end{abstract}

\vspace{10pt}

\end{centering} 

\vspace{125pt}

\end{titlepage}

\setcounter{footnote}{0}
\section{Introduction}
\label{sec:Introduction}
In our previous paper \cite{GovaertsEtAl2007}, J Phys A \textbf{40} (2007) 12095--12111 (hereinafter referred to as I), we initiated an investigation into worldline formulations of elementary systems possessing the extended symmetries of reciprocal relativity. These make up the so-called  quaplectic\footnote{See S. G Low, {e-print {\tt arXiv:math-ph/0502018}} for an explanation of this name.} group $Q(D\!-\!1,1)\cong U(D\!-\!1,1)\ltimes H(D)$, the
semi-direct product of the pseudo-unitary group with the Weyl-Heisenberg group in $D$ dimensions.
The latter is realised as the central extension of a $2D\!+\!1$-dimensional translation group, in accord with the original vision of Born \cite{Born1949,Born1949a,Green1949,Green1949a} whereby `position' and `momentum' are reciprocally equivalent, and generalised in recent work by Low \cite{Low2002,Low2005a,Low2005b} to encompass the full $(D\!+\!1)^2$-dimensional quaplectic transformation group. The novel feature in our formulation was a compact `phase' coordinate $\theta(\tau)$, in additional to worldline `position' and `momentum' coordinates $x^\mu(\tau)$ and $p_\mu(\tau)$; the conserved $\theta$-momentum thus introduces a quantum number with a discrete spectrum, and sets the scale of Planck's constant in the Heisenberg algebra (and can be regarded as a superselected quantity).

The worldline model presented in I is defined by the action (in $D$-dimensional Minkowski space) 
\begin{align}
\label{eq:WorldlineAction}
S =& \,  \int d\tau \big( \frac{1}{e(\tau)} L + \Lambda e(\tau) \big), \nonumber \\
L = & \, - \frac 12 \left( \frac{dx^\mu}{d\tau} \frac{dx_\mu}{d\tau} +
\kappa_0^2 \frac{dp_\nu}{d\tau} \frac{dp^\nu}{d\tau} \right) +
{\texttt{C}}\frac{\kappa_0}{\lambda_0}\left( \frac{d\theta}{d\tau} -
\lambda_0 \left(x^\mu(\tau) \frac{dp_\mu}{d\tau} -
p_\nu(\tau)\frac{dx^\mu}{d\tau} \right)\right)^2,
\end{align}
where $e(\tau)$ is the worldline einbein, $\kappa_0 = c/b$, where  $b$ is Low's parameter of maximum force (rate of change of momentum), $c$ is the speed of light, $\lambda_0$ has units of action${}^{-1}$, and ${\texttt C}$ is an arbitrary numerical constant. In I, a complete analysis of Noether charges and constraints arising from the symmetries of (\ref{eq:WorldlineAction}) was presented (for the methodology of constraint quantisation see for example \cite{Govbook}). The state space of the system (before imposition of the constraint) corresponds to the direct product of \emph{two} independent $D$-dimensional Heisenberg algebras -- one generated by the conserved generators ${\mathcal X}^\mu$, ${\mathcal P}_\nu$ of translations in the original `position' and `momentum' worldline coordinates, and a second auxiliary, non-conserved set ${\mathbb X}^\mu$, ${\mathbb P}_\nu$ (both sets have the same central extension, the conserved $\theta$-momentum, $\Pi_\theta$). Imposition of the first class constraint, the generator of local time reparametrisations, on physical
state space enforces identification \cite{Gov2} of the world-line cosmological constant $\Lambda$ in the model, with a fixed value of the quadratic Casimir of the quaplectic symmetry group.  However, both sets of Heisenberg generators provide the material for the construction of the (conserved) generators of the homogeneous $U(D\!-\!1,1)$ component of the quaplectic algebra,
\begin{align}
\label{eq:EasZZ}
E_{\mu \nu} := & \, \frac 12 \{\overline{\mathcal Z}_\mu, {\mathcal Z}_\nu \}
                       + \frac 12 \{\overline{\mathbb Z}_\mu, {\mathbb Z}_\nu \} 
\end{align}
in a complex basis with covariant combinations
\begin{align}
\label{eq:ComplexZ}
{\mathcal Z}_\mu = & \, \frac{1}{\sqrt{2}}(-{\mathcal X}_\mu + i{\mathcal P}_\nu),
\qquad 
{\mathbb Z}_\mu = \frac{1}{\sqrt{2}}(-{\mathbb X}_\mu + i{\mathbb P}_\nu),
\end{align}
and wherein the generators of Lorentz transformations are, for example, the total orbital-type combinations $({\mathcal X}_\mu {\mathcal P}_\nu -{\mathcal X}_\nu {\mathcal P}_\mu) +
({\mathbb X}_\mu {\mathbb P}_\nu -{\mathbb X}_\nu {\mathbb P}_\mu)$. Defining $U = {E^\mu}_\mu$, the $Q(D\!-\!1,1)$ Casimir operator reads
\begin{align}
\label{eq:CasimirUs}
C_1  = & \,  U \Pi_\theta -  \textstyle{\frac 12} \big( {\mathcal P}^\mu {\mathcal P}_\mu  + {\mathcal X}^\mu {\mathcal X}_\mu \big) 
\equiv  
\textstyle{\frac 12} \big( {\mathbb P}^\mu {\mathbb P}_\mu  + {\mathbb X}^\mu {\mathbb X}_\mu \big),  
\end{align}
which is immediately recognisable as an ingredient of the first class constraint, to be imposed on physical state space as an operator condition,
\begin{align}
\label{eq:Constraint}
\textstyle{\frac 12} \big( {\mathbb P}^\mu {\mathbb P}_\mu  + {\mathbb X}^\mu {\mathbb X}_\mu \big) +  \textstyle {\frac 12} \Pi_\theta^2 = & \,  \Lambda. 
\end{align}
Thus for a nontrivial solution, the worldline cosmological constant must lie in the spectrum of the quaplectic Casimir, modulo the square of the conserved $\theta$-momentum\footnote{See I for details of the correct way to scale out physical constants to produce the dimensionless degrees of freedom assumed in the present treatment. For example, the correct definition of the Heisenberg algebra generators depends on the sign of the $\Pi_\theta$  eigenvalue (assuming this to be nonvanishing).}.

In I, the auxiliary operators ${\mathbb X}^\mu$, ${\mathbb P}_\nu$ were treated as `covariant oscillators', with a Lorentz invariant zero mode state, but acting on a space containing some negative-norm states, associated with the fact that the indefinite sign of the Minkowski metric necessarily leads to at least one set of raising and lowering operators with a `wrong-sign' commutation relation. In this case the only consistent solution is to arrange things so that the cosmological constant balances the contribution to the constraint from $\Pi_\theta$, and only the vacuum state itself survives; the auxiliary space collapses, and only spinless fields occur in the spectrum. However, because there are no other restrictions on the physical energy-momentum vector ${\mathcal P}_\nu$, its eigenvalues $p_\nu$ will produce a continuous range of values of $p\!\cdot\!p$, including not only massive and massless, but also unphysical\footnote{In the following, a distinction is intended between `physical state space', meaning the state space after imposition of the constraint, and `physical and unphysical particles', meaning that, as well as standard particle states, physical state space also contains unconventional, and hence unphysical in that sense, particle content.}, tachyonic (spacelike) branches. We refer the reader to I for a full discussion of the model.

In this present paper, the auxiliary space is treated using standard oscillators, with correct-sign commutation relations. As a result, only positive-norm states occur (the auxiliary space is therefore isomorphic to a product of standard $L^2({\mathbb R})$ spaces). Compared with the covariant approach however, the zero-mode state in the auxiliary space is not Lorentz- or Born reciprocal-invariant, but takes its place in the weight decompositions of the appropriate extended little algebras. The auxiliary `spin'-containing space is vastly enriched as compared with the covariant approach, and notwithstanding the continuous range of $p\!\cdot\!p$, it is instructive carry out a careful case-by-case analysis of the spin content. 

It is this task which we are at pains to present in detail here, with the  
restriction to $D=4$ dimensions, so that the full spectrum of states
in the physical state space after imposition of the constraint, with both physical and `unphysical' particle content, can be compared with the standard classification of unitary irreducible representations of the Poincar\'{e} group \cite{Wigner1939}.

As already mentioned, the physical worldline model defined by (\ref{eq:WorldlineAction}) is developed in I, to which the reader is referred for details. In \S 2 below, we commence analysis of the problem of identifying the appropriate little algebras, and accompanying degeneracy algebras, for each of the orbit classes $p\!\cdot\!p >0$, $<0$, $\equiv 0$, and $=0$ (massive, spacelike (tachyonic), null and massless, respectively). We take as read the details of the quantisation of the model carried out in I; for completeness however, we include in the appendix \S \ref{subsec:Q31}, a discussion of the structure of the quaplectic Lie algebra and bases relevant for physics. The main technical details of the derivation of 
group branching rules are held over to \S \ref{subsec:Sp2Unirreps} (unitary irreducible representations of $Sp(2,{\mathbb R})$ and products of discrete series representations), \S \ref{subsec:Sp2dDuals} (dual subalgebras of $Sp(8,{\mathbb R})$ and the $>0$, $<0$, $\equiv 0$ cases), \S \ref{subsec:Massive} (massive case), \S \ref{subsec:Spacelike} (tachyonic, spacelike case), \S \ref{subsec:Null} (null case), and finally \S \ref{subsec:E2Cases} (dual $E(2)$ subalgebras and massless states).

Results are collected in summary form as tables
\ref{tab:LittleAlgTab} and \ref{tab:Results}.
The analysis of representations found is carried out in the concluding \S 3, with attention to how the constraint selects the physical state space in each case.  
The overview of results is complemented by a discussion of potential future directions and refinements.
\section{Wigner method for extended worldline system}
\label{sec:Wigner}
As must be the case for valid quantisation of any system admitting classical symmetries, the corresponding state space carries appropriate unitary representations of the symmetry group in question. In the case of the worldline system (\ref{eq:WorldlineAction}), this is of course the quaplectic Lie group $Q(3,1)\cong U(3,1)\ltimes H(4)$. 

The structure of the corresponding Lie algebra is given in the appendix, \S  \ref{subsec:Q31}. As is clear from (\ref{eq:EasZZ}) above, the homogeneous generators ${E^{\mu}}_{\nu}$ of $U(3,1)$ are constructed using the material provided by both independent 4 dimensional Heisenberg algebras (generated by ${\mathcal Z}_\mu$, $\overline{\mathcal Z}_\nu$, as well as the auxiliary set ${\mathbb Z}_\mu$, $\overline{\mathbb Z}_\nu$); whereas \emph{the} physical Weyl-Heisenberg group is generated by the set ${\mathcal Z}_\mu$, $\overline{\mathcal Z}_\nu$, associated with the conserved Noether charges. However, from (\ref{eq:CasimirUs}) it turns out that the $Q(3,1)$ Casimir invariant $C_1$ depends only on the auxiliary Heisenberg generators. In fact from (\ref{eq:EasZZ}) and the construction (\ref{eq:SpinOrbitE}), it is clear that the derived quantities ${e^\mu}_\nu$, 
which are generators of $U(3,1)$ but which \emph{commute} with the physical Heisenberg algebra, are precisely
$\frac 12 \{\overline{\mathbb Z}^\mu,{\mathbb Z}_\nu \}$; the $Q(3,1)$ Casimir invariants are traces of matrix products of ${e^\mu}_\nu$ (see \S \ref{subsec:Q31}) and
in this realisation of the of $U(3,1)$ Lie algebra, it is known \cite{Shamaly1974} that the linear one is the only independent Casimir invariant. The role of the auxiliary space carrying representations of $U(3,1)\supset O(3,1)$ via the ${e^\mu}_\nu$ is brought out for instance by the expression for the generators of Lorentz transformations (see (\ref{eq:EasZZ}), (\ref{eq:ExtendedQplecticCR})), namely
\begin{align}
\label{eq:SpinOrbitL}
L_{\mu \nu} = & \, i({\mathcal X}_\mu {\mathcal P}_\nu- {\mathcal X}_\nu {\mathcal P}_\mu) + i({\mathbb X}_\mu {\mathbb P}_\nu - {\mathbb X}_\nu {\mathbb P}_\mu) \equiv 
i({\mathcal X}_\mu {\mathcal P}_\nu- {\mathcal X}_\nu {\mathcal P}_\mu) + {\mathbb L}_{\mu \nu}
\end{align}
which means that ${\mathbb L}_{\mu \nu}$ is identified with spin.

With a view to its reduction with respect to unitary irreducible representations of the space-time Poincar\'{e} algebra, it is possible to diagonalise the physical four-momentum ${\mathcal P}_\mu$ (adopting the usual quantum-mechanical Schr\"{o}dinger representation with ${\mathcal P}_\mu \rightarrow -i\hbar \partial/\partial x^\mu$ for suitable functions on Minkowski space), but for the auxiliary space, for enumerative purposes, to identify the auxiliary 4-dimensional Heisenberg algebra with 4 pairs of standard oscillator raising and lowering operators, via (\ref{eq:ZasOsc}). With the standard commutation relations 
\begin{align}
{[}a, a^\dagger {]} = & \, 1, \qquad {[}b_i, b_j^\dagger {]} = \delta_{ij}, \quad i,j = 1,2,3, \nonumber
\end{align}
and the usual introduction of the zero mode state annihilated by the lowering operators, a suitable basis for the auxiliary space is therefore provided by the number states $|n_0, n_1,n_2,n_3 \rangle$ (see (\ref{eq:ZeroOccupancy})), $n_0, n_i = 0,1,\cdots$ being the eigenvalues of the respective number operators $\widehat{N}_0 = a^\dagger a$, $\widehat{N}_i = b_i^\dagger b_i$, $i=1,2,3$. 

Finally the full quantum space of states (before imposition of the constraint) is spanned by the basis\footnote{A further direct product with a one-dimensional space associated with the fixed $\theta$-momentum has been omitted for the sake of clarity; see (\ref{eq:NumberConstraint}) below.} 
\begin{align}
\label{eq:NumberStates}
|p^\mu \rangle \otimes | n_0,n_1,n_2,n_3 \rangle \equiv & \, |p^\mu; n_0,n_1,n_2,n_3 \rangle .
\end{align}
Now, from the general expression for the Casimir invariant (\ref{eq:QuadraticCasimirAll}), and its form (\ref{eq:CasimirUs}) in the present realisation, the constraint (\ref{eq:Constraint}) to be imposed on physical states amounts to the projection of the generic number states $|p^\mu; n_0,n_1,n_2,n_3 \rangle$ on to a certain selected eigenspace of the covariant oscillator hamiltonian,
$\frac 12({\mathbb P}\!\cdot \!{\mathbb P} + {\mathbb X}\!\cdot \!{\mathbb X})$. For consistency with the identifications to be made below between this operator and various key symmetry generators, we adopt the following form of the constraint:
\begin{align}
\label{eq:NumberConstraint}
\textstyle{\frac 12}( n_1+n_2+n_3 - n_0) + \textstyle{\frac 12} = & \, \Delta, \quad \mbox{where} \quad 
 \Delta := \textstyle{\frac 12}\Lambda -\textstyle{\frac 14} (n_\theta+\sigma)^2,
\end{align} 
where we assume that a fixed eigenvalue $n_\theta+\sigma$ of the compact $\theta$-momentum ${\Pi}_\theta$ has been selected, for some integral $n_\theta$, up to a modular parameter $0\le \sigma < 1$ (see \cite{Gov3} and I).

Since $p^\mu \in {\mathbb R}^4$, there is \emph{no} restriction on the energy-momentum squared, $p\!\cdot\!p$. In analysing the particle content of physical state space by carrying out the complete decomposition of the (irreducible) unitary representations of $Q(3,1)$ specified by the states (\ref{eq:NumberStates}) subject to (\ref{eq:NumberConstraint}), with respect to the Poincar\'{e} group, the spectrum in the massive and tachyonic (spacelike) branches will therefore be a continuum, with $\infty > p\!\cdot\!p >0$, and $-\infty < p\!\cdot\!p<0$ respectively. Moreover, we will see below that, for both the massless and null cases where $p\!\cdot\!p \equiv 0$, the reduction turns out also to be in the form of a direct integral. 

From (\ref{eq:SpinOrbitL}), it is necessary to refine the physical state space analysis by diagonalising the \emph{second} Poincar\'{e} group Casimir, corresponding to the spin quantum number. For this task we take up the traditional method of Wigner \cite{Wigner1939}: for each orbit class, it is sufficient to work with 4-momentum fixed in some standard frame, $\!\pcirc\!$, and give the complete reduction of state space with respect to the appropriate little algebra (the subalgebra of the Poincar\'{e} Lie algebra which fixes the given standard 4-momentum). In the present realisation, the physical little algebra, say ${\mathcal L}(\!\pcirc\!)$, is extended by a dual or commutant ${\mathbb L}(\!\pcirc\!)$ within $U({\mathbb H}_4)$. That is, there is a certain subalgebra of the full enveloping algebra of the auxiliary Heisenberg algebra, which controls the degeneracy of little algebra unirreps, subject to the constraint (\ref{eq:NumberConstraint}) which as we have seen, projects the physical state space onto a unitary irreducible representation of the  full quaplectic algebra.

In the massive, spacelike and null cases, the little algebras, denoted ${\mathcal L}^{(>0)}$, ${\mathcal L}^{(<0)}$ and ${\mathcal L}^{(\equiv 0)}$, are the Lie algebras of the orthogonal groups $SO(3)$, $SO(2,1)$ and $SO(3,1)$ respectively. A natural way to find their commutants ${\mathbb L}(\!\pcirc\!)$ follows if the number space of the auxiliary oscillator modes is identified with an appropriate (metaplectic) unitary representation of the Lie algebra of the eight-dimensional symplectic group $Sp(8,{\mathbb R})$, using the general embeddings based on the group branchings $Sp(2d, {\mathbb R}) \supset Sp(2,{\mathbb R}) \times SO(d)$ or $Sp(2d, {\mathbb R}) \supset Sp(2,{\mathbb R}) \times SO(d-1,1)$ for the appropriate $d$ and real forms.

In the null case for example, $\!\pcirc = (0,0,0,0)$ and the the little algebra is the entire Lorentz group Lie algebra ${\mathcal L}^{(\equiv 0)} =SO(3,1)$; the noncompact embedding with $d=4$ applies, and the commutant ${\mathbb L}^{(\equiv 0)}$ is denoted $Sp^{(0123)}(2,{\mathbb R})$ reflecting that it is the diagonal sum of the \emph{four} $Sp(2,{\mathbb R})$ oscillator algebras, one for each Cartesian direction in Minkowski space (for details of unitary irreducible representations of $Sp(2,{\mathbb R})$ in relation to oscillator representations, and rules for the reduction of direct products of such representations, see \S \ref{subsec:Sp2Unirreps}). On the other hand for $\!\pcirc = (0,0,0,\!\pcirc\!\!\!\!{}_z)$ (the spacelike case), the little algebra $SO(2,1)$, generated by three-dimensional Lorentz transformations in directions 0,1,2 in Minkowski space, clearly commutes with both the diagonal $Sp^{(012)}(2,{\mathbb R})$ as well as $Sp^{(3)}(2,{\mathbb R})$, so 
${\mathcal L}^{(<0)} +{\mathbb L}^{(<0)} = SO(2,1) + \big(Sp^{(012)}(2,{\mathbb R})+Sp^{(3)}(2,{\mathbb R})\big)$. Similar considerations show that
for $\!\pcirc = (mc,0,0,0)$ (the massive case),
${\mathcal L}^{(>0)} +{\mathbb L}^{(>0)} = SO(3) + \big(Sp^{(0)}(2,{\mathbb R})+Sp^{(123)}(2,{\mathbb R})\big)$. Finally in the massless case, the appropriate commutant is to be found within $U({\mathbb H}_4)$ itself rather than $Sp(8,{\mathbb R})$, and we find 
${\mathcal L}^{(=0)} + {\mathbb L}^{(=0)} = {\mathcal E}(2) + {\mathbb E}(2)$, a direct sum of two three-dimensional Euclidean Lie algebras.

Results are listed case-by-case in tables \ref{tab:LittleAlgTab} and \ref{tab:Results}. 
Table \ref{tab:LittleAlgTab} 
provides the standard reference 4-momenta $\!\pcirc\!$, the corresponding little algebra, and the commutant in each case. Also given in each case is the explicit combination of diagonal generators of the dual algebra which represent the constraint operator, the difference of number operators $\frac 12(\widehat{N}_1 + \widehat{N}_2 + \widehat{N}_3 - \widehat{N}_0) + \frac 12$ (whose $\Delta$-eigenspace, from (\ref{eq:NumberConstraint}) provides the physical state space). Details of the embeddings within $Sp(8,{\mathbb R})$ and the general $Sp(2d, {\mathbb R}) \supset Sp(2,{\mathbb R}) \times SO(d)$ or $Sp(2d, {\mathbb R}) \supset Sp(2,{\mathbb R}) \times SO(d-1,1)$ constructions are provided in \S \ref{subsec:Sp2dDuals}. The identification 
of the dual algebra for the massless case is described in \S \ref{subsec:E2Cases}. In \S \S \ref{subsec:Sp2dDuals} - 
\ref{subsec:E2Cases} can be found details of the unitary irreducible representations of the occurring compact and non-compact Lie algebras in addition to $Sp(2, {\mathbb R})$ (which is treated in \S \ref{subsec:Sp2Unirreps}). In this notation, table \ref{tab:Results} lists for each orbit class, the physical state space(s)
occurring, and also their degeneracy, in terms of unitary irreducible representations of the respective dual, commutant algebras (projected onto the $\Delta$-eigenspace as indicated in table \ref{tab:LittleAlgTab}).

\section{Results and discussion}
We begin by stating the results of the analysis sketched in the foregoing, as summarised in tables \ref{tab:LittleAlgTab} and \ref{tab:Results}, for each sector of the $p\!\cdot\!p$ spectrum, and paying attention to the constraint projection. \\[.3cm]
\textbf{Massive states}:\\
For fixed $p\!\cdot\!p = m^2c^2 >0$, there is an infinite series of integer-spin particles $\ell = 0, 1, 2, \cdots$, with for each such spin, a degeneracy governed by the tensor product with the indicated projective representation of the dual algebra, which can be enumerated as follows. States of the \emph{reducible} representation $D^-_{-1/4} \!\oplus \! D^-_{-3/4}$ are spanned by the eigenvectors of $K_0^{(0)}$, with spectrum $-\frac 14 -\frac 12 k^{(0)}$, $k^{(0)}= 0,1,2,\cdots$, whereas for fixed $\ell$, we have $K_0^{(123)}$ eigenvalues of the form $\frac 12 \ell + \frac 14 + k^{(123)}$, $k^{(123)}=  0,1,2,\cdots$. Thus the constraint reads
\begin{align}
\label{eq:MassiveConstraint}
(\textstyle{\frac 12} \ell + \textstyle{\frac 14} + k^{(123)})+(-\textstyle{\frac 14} -\textstyle{\frac 12} k^{(0)}) = & \, \Delta, \quad \mbox{or} \quad
k^{(123)} -\textstyle{\frac 12} k^{(0)} = \Delta -\textstyle{\frac 12}\ell 
\end{align}
so that an infinite tower of recurrences of each $\ell$ exists (depending on whether $\Delta -\textstyle{\frac 12}\ell$ is integral or half-integral, these are associated with even or odd modes of the ${}^{(0)}$ oscillator). In fact, regarding the $K_0^{(0)}$ eigenvalue, or just $k^{(0)}$, as being fixed by the choice of $k^{(123)} = 0,1,2,\cdots$, a weight diagram of ${[}\ell{]}$ versus $k^{(123)}$ just corresponds to the orbital angular momentum content of a three-dimensional isotropic simple harmonic oscillator system. \hfill $\Box$ \\
\textbf{Spacelike (tachyonic) states}:\\
For fixed $p\!\cdot\!p <0$, the allowed unirreps of $SO(2,1)$ (consistent with the constraint) are listed in table \ref{tab:Results}. The analysis is complicated in this case by three different cases of discrete series representations, and also a continuous series contribution.  
For example, the auxiliary representation spaces tensored either with the $D^+_\ell$ \emph{or} $D^-_\ell$ discrete $SO(2,1)$ unirreps (for $\ell = 1,2,\cdots$), are spanned by the eigenstates of $K_0^{(012)}$, with eigenvalues $\frac 12 \ell -\frac 14 + k^{(012)}$, $k^{(012)} = 0,1,2\cdots$; together with eigenstates of $K_0^{(3)}$, with eigenvalues
$\frac 14 + \frac 12 k^{(3)}$, $k^{(3)} = 0,1,2\cdots$, and the constraint reads
\begin{align}
\label{eq:TachyonicConstraintA}
(\textstyle{\frac 12} \ell - \textstyle{\frac 14} + k^{(012)})+(\frac 14 + \frac 12 k^{(3)}) = & \, \Delta, \quad \mbox{or} \quad
k^{(012)} + \textstyle{\frac 12} k^{(3)} =  \Delta - \textstyle{\frac 12} \ell, \,\, \ell = 1,2,3 \cdots.          
\end{align}
The analysis of the remaining cases proceeds similarly. The singlet representation ${[}0{]}$ of $SO(2,1)$ is accompanied by discrete series representations $D^\pm_{-1/4}$ of $Sp^{(012)}(2,{\mathbb R})$ and the constraint becomes
\begin{align}
D^+_{-1/4}: \qquad \,\,\,\, (  \textstyle{\frac 14} + k^{(012)})+(\frac 14 + \frac 12 k^{(3)}) = & \, \Delta, \quad \mbox{or} \quad
k^{(012)} + \textstyle{\frac 12} k^{(3)} =   \Delta -\textstyle{\frac 12}, \nonumber \\
\mbox{and} \qquad D^-_{-1/4}: \qquad
( -\textstyle{\frac 14} - k^{(012)})+(\frac 14 + \frac 12 k^{(3)}) = & \, \Delta, \quad \mbox{or} \quad
-k^{(012)} + \textstyle{\frac 12} k^{(3)} =   \Delta ,
\label{eq:TachyonicConstraintB}
\end{align}
respectively. By contrast, the continuous series representation $D(-\frac 12 +is)$ occurs as a direct integral over $s$, tensored with the corresponding unirrep $D(-\frac 12 +2is)$ of $Sp^{(012)}(2,{\mathbb R})$ (which is doubly degenerate). The spectrum of $K_0^{(012)}$ is $\frac 12 m^{(012)} + \frac 14$ for \emph{integer} $m^{(012)}$, so that together with the $K_0^{(3)}$ eigenstates the  constraint condition reads
\begin{align}
\label{eq:TachyonicConstraintC}
(\textstyle{\frac 12}m^{(012)} +\textstyle{\frac 14})+(\frac 14 + \textstyle{\frac 12} k^{(3)}) = & \, \Delta, \quad \mbox{or} \quad
\textstyle{\frac 12}m^{(012)} + \textstyle{\frac 12} k^{(3)} =   \Delta - \textstyle{\frac 12}.
\end{align}
See below for a discussion of the issue of which of these tachyonic states can survive the constraint projection. \hfill $\Box$ \\ \\
\textbf{Null states}:\\
The analysis proceeds similarly to the discussion of the tachyonic states above. Taking into account the spectrum of $K^{(0123)}$ in the two different cases (discrete series or (doubly degenerate) continuous principal series unirreps of $SO(3,1)$) we have
\begin{align}
k^{(0123)} = & \,  \Delta - \textstyle{\frac 12}\ell -\textstyle{\frac 12}, \quad \ell = 0,1,2,\cdots \nonumber \\
\mbox{or}\qquad 
\qquad \textstyle{\frac 12} m^{(0123)} = & \,  \Delta - \textstyle{\frac 12},  
\label{eq:NullConstraint}
\end{align}
respectively.\hfill $\Box$ \\ 
\textbf{Massless states}:\\
Unitary irreducible representations for massless states have a decomposition over continuous ${\mathcal E}(2)$ series unirreps, for Euclidean momentum $0<\pi<\infty$ (with arbitrary helicity $0, \pm 1, \pm 2, \cdots$ ) while table \ref{tab:Results} shows that for fixed $\pi$, there is (up to re-scaling of the dual momentum label) a \emph{further} continuum of unirreps of the analogous ${\mathbb E}(2)$ dual algebra. Within each such dual unirrep, the constraint simply selects the eigenvalue $\Delta$ of the diagonal generator. \hfill $\Box$ \\

As mentioned in the introduction, and as is clear from the above discussion and table \ref{tab:Results}, the present model shows a complex structure of `physical' particle states. Firstly,   there is no selected mass scale, as the spectrum of $p\!\cdot\!p$ is over the whole real line. Even for the massive branch $p\!\cdot\!p = m^2c^2>0$, there is an infinite number of particles with spins $\ell = 0,1,2,\cdots$, each of which is countably degenerate, so that the spin content as a whole is equivalent to the orbital angular momentum decomposition of a nonrelativistic isotropic three-dimensional simple harmonic oscillator system. As to the $p\!\cdot\!p =0$ branch, unfortunately, in this model, the massless `particle' states are not of the conventional helicity type, but rather can have nonzero Euclidean momentum (and hence arbitrary integer helicity), with each such non-minimal massless unirrep itself being continuously degenerate.

All of the tachyonic states identified constitute an unacceptable violation of causality, but for completeness we have listed them in full as they are unavoidable consequences of our present construction. However, it is apparent from (\ref{eq:TachyonicConstraintA}) - (\ref{eq:TachyonicConstraintC}) that, remarkably, it is possible simply by a careful choice of $\Delta$ to eliminate some of these states. For example, $\Delta =0$ or $\Delta = \frac 12$ removes solutions of (\ref{eq:TachyonicConstraintA}) and (\ref{eq:TachyonicConstraintB}a), while in (\ref{eq:TachyonicConstraintB}b) and (\ref{eq:TachyonicConstraintC}) the parity of $k^{(3)}$ and $k^{(012)}$ or $m^{(012)}$ in the respective oscillator spaces (whether even or odd occupation numbers are admissible) is correlated according to whether $\Delta$ is integral or half-odd integral. Likewise, the null states, with $\!\pcirc \equiv 0$, are also not associated with conventional particle states, but similar comments about the possibility of at least partial elimination of some of these states apply - for example from (\ref{eq:NullConstraint}a), it is evident that again a choice such as $\Delta = 0$ or $\frac 12$ would serve to remove these unirreps (while leaving the contribution from the continuous series representations unaffected). 

We conclude this discussion with some comments on extensions of the present work, which may allow some of these issues arising from the specific results, to be addressed (related comments were made in the concluding remarks of I, but in the context of the results of that earlier approach).

It is striking that,
despite the presence of dimensionful constants $\kappa_0$ and $\lambda_0$ ($b/c$, and $\hbar$) in the worldline action, there is no selected mass scale. However, from the geometrical point of view, given the close relationship between semi-direct product groups and coset spaces, it is natural to expect that  further generalisations of the action (\ref{eq:WorldlineAction}) can be constructed. Such a modification of the geometry of the `coordinates' $x^\mu(\tau), p_\nu(\tau), \theta(\tau)$ may suggest that the continuous spectrum of $p\!\cdot\!p$ in the present model is at least an artefact of having a `flat' target space.  

It is clear from, say, (\ref{eq:SpinOrbitL}) and confirmed by table \ref{tab:Results}, that only integer spin (and helicity) states are possible in the present construction. A natural further step would thus be towards the equivalent of spinning particle or superparticle versions of reciprocally invariant worldline systems. It is conceivable, although not evident how at present, that in this context, the combination of first class constraints in such a super-formulation, might indeed allow a projection onto acceptable, \emph{conventional} types of particles in the physical state space, and exorcise the non-minimal and `unphysical', in a conventional sense, particle states allowed in our present construction.

Finally there is the question of the interpretation of unconventional unirreps such as the continuous Euclidean momentum, infinite helicity massless states (see table \ref{tab:Results}). If they survive in a complete and otherwise consistent model, then they deserve to be taken seriously as a potential component of `exotic matter'. Presumably, the mass or energy scale of the Born constant $b$ (for example equating the `Born mass'  $\sqrt{\hbar b/c^3}$ with the Planck mass $\sqrt{\hbar c/G_N}$ yields $b \cong c^4/G_N$ $\cong 10^{44}\, N$) should be invoked to suppress couplings of such nonstandard states to conventional matter; they would then tend to be invisible to conventional detectors made of normal particles\footnote{See \cite{JarvisMorgan2006} for comments on such issues as the indeterminacy principle in the context of reciprocity; a local gauged version of quaplectic symmetry has been considered in 
\cite{Castro2008}.}.

We leave such speculations to future work. Born reciprocity is evidently an original and challenging starting-point for a theory of generalised elementary particles, which deserves serious study in the same vein as other higher-dimensional, or string-like models. It shares with them some of the same problems, such as being beset by unphysical particle spectra, and an explosion of modes, although it is unique in that it does not involve any additional `higher dimensions' (other than the compact $\theta$-coordinate, whose conserved momentum is anyway assumed to be superselected). Of course, it goes without saying that it runs foul of the famous `no go' theorems \cite{ORaif} which forbid extensions of Poincar\'{e} invariance in space-time in the context of the usual postulates of local quantum field theory, so the unrealistic particle spectrum in the present worldline model is not unanticipated. But, at base, reciprocal relativity is radically different from standard relativistic physics. Not least, it postulates a maximum momentum transfer rate, $b$, between interacting systems -- analogously to the maximum displacement rate -- the speed of light, $c$, in relativity. More fundamentally, it overturns conventional understandings of kinematics and dynamics in its very denial of the existence of inertial frames, and its insistence on the ubiquity of interactions, even in the nonrelativistic regime \cite{Low2008}.

\subsection*{Acknowledgements}
S. O. M. acknowledges financial support from a Commonwealth Postgraduate Award.
We thank Jan Govaerts, Stephen Low, Robert Delbourgo and Rutwig Campoamor-Stursberg for fruitful discussions during the course of this work.

\newpage  
\begin{appendix}
\setcounter{equation}{0}
\renewcommand{\theequation}{{A}-\arabic{equation}}
\section{Appendix}
\label{sec:appendix}
\subsection{Quaplectic algebra $Q(3,1)$}
\label{subsec:Q31}
The 25-dimensional quaplectic Lie algebra in $D=4$ dimensions is generated by ${E^\mu}_\nu$, $\mu,\nu = 0,1,2,3$ such that (in unitary representations)
\begin{align}
\label{eq:HermiticityU31}
({E^\mu}_\nu)^\dagger = & \,  \eta^{\mu \rho} \eta_{\nu \sigma} 
({E^\sigma}_\rho)
\end{align}
which generate the real Lie algebra of $U(3,1)$,
\begin{align}
\label{eq:CommutatorsU31}
{[} {E^\mu}_\nu, {E^\rho}_\sigma {]} = & \, {\delta^\mu}_\sigma {E^\rho}_\nu - {\delta_\nu}^\rho 
{E^\mu}_\sigma,
\end{align}
together with the complex vector operator $Z^\mu$ and its conjugate 
${\overline{Z}}\mbox{}^\mu$,
\begin{align}
\label{eq:Hermiticity Z}
(Z^\mu)^\dagger = & \, \eta^{\mu \rho} {\overline{Z}}\mbox{}_\rho \equiv 
{\overline{Z}}\mbox{}^\mu
\end{align}
which fulfil the Heisenberg algebra (with central generator $I$)
\begin{align}
\label{eq:HeisenbergZ}
{[} Z^\mu, {\overline{Z}}\mbox{}^\nu {]} = & \, -   
\eta^{\mu \nu} I,   \\
{[} Z^\mu, Z^\nu {]} = &\, 0 \, = {[} {\overline{Z}}\mbox{}^\mu, 
{\overline{Z}}\mbox{}^\nu {]}.
\end{align}
The $E$ and $Z$ satisfy the commutation relations:
\begin{align}
\label{eq:CommutatorsEZ}
{[} {E^\mu}_\nu, {\overline{Z}}\mbox{}^\rho {]} = & \, {\delta_\nu}^\rho 
{\overline{Z}}\mbox{}^\mu, \nonumber \\
{[} {E^\mu}_\nu,  Z^\rho {]} = & \, -  \eta^{\mu \rho}  Z_\nu.
\end{align}
In the above, the Lorentz metric $\eta_{4 \times 4} = diag(+1,-1,-1,-1)$ 
is adopted together with standard conventions for raising and lowering 
indices.

Relativistic position and momentum operators $X^\mu$, $P^\nu$ are 
defined as the quadrature components of $Z^\mu$ and 
${\overline{Z}}\mbox{}^\mu$, namely\footnote{See footnote 2.}
\begin{align}
\label{eq:Quadratures}
Z^\mu =   \textstyle{\frac{1}{\sqrt{2} }}( -{X^\mu}{} + i 
 {P^\mu}{}), & \, \quad
\overline{Z}^\mu =   \textstyle{\frac{1}{\sqrt{2} }}(-{X^\mu}{} - i 
 {P^\mu}{}), \nonumber \\
 {X^\mu}{} =   \textstyle{\frac{1}{\sqrt{2} }}(Z^\mu + 
{\overline{Z}}\mbox{}^\mu), &\, \quad
 {P^\mu}{} =   \textstyle{\frac{i}{\sqrt{2} }}(Z^\mu - 
{\overline{Z}}\mbox{}^\mu), \nonumber 
\\                                                                               
\mbox{with} \qquad {[}  {X^\mu}{}, 
 {P_\nu}{}{]} =   
i   {\delta^\mu}_\nu,  \quad \mbox{and} & \, \quad
{[} X^\mu, X^\nu {]} = 0 = {[}  P^\mu, P^\nu {]}.
\end{align}

The structure of the quaplectic algebra is made more transparent in 
terms of auxiliary generators ${\{} {e^\mu}_\nu {\}}$ which provide a 
`spin-orbit' like decomposition,
\begin{align}
\label{eq:SpinOrbitE}
{e^\mu}_\nu :=  {E^\mu}_\nu - \frac{1}{2 
I}{\{}{\overline{Z}}\mbox{}^\mu , Z_\nu{\}}, \quad
& \,
{E^\mu}_\nu =   {e^\mu}_\nu +  \frac{1}{2 
I}{\{}{\overline{Z}}\mbox{}^\mu , Z_\nu {\}},
\end{align}
such that the ${e^\mu}_\nu$ satisfy the $U(3,1)$ algebra, but commute 
with $H(4)$:
\begin{align}
\label{eq:EeCommutators}
{[} {E^\mu}_\nu, {e^\rho}_\sigma {]} = & \, {\delta^\mu}_\sigma {e^\rho}_\nu - {\delta_\nu 
}^\rho{e^\mu}_\sigma ,  \nonumber \\
{[} {e^\mu}_\nu, {e^\rho}_\sigma {]} = & \,  {\delta^\mu}_\sigma {e^\rho}_\nu -  {\delta_\nu 
}^\rho{e^\mu}_\sigma , \nonumber \\
  {[} {e^\mu}_\nu, Z^\rho {]} = &\, 0 =  {[} {e^\mu}_\nu, 
\overline{Z}^\rho {]}.
\end{align}
As confirmed by the details of Mackey induced representation theory applied to this case, (see 
\cite{Low2005a}), it is clear from this sketch that a generic unitary irreducible 
representation (unirrep) of the quaplectic group can be associated with 
the tensor product of a unirrep of $U(3,1)$ (provided by nonzero 
${e^\mu}_\nu$), with suitable unirrep(s) of the Weyl-Heisenberg algebra 
$H(4)$. The latter can of course themselves be identified via induced 
representations \cite{Wolf1975}.

Finally, in relation to the general quaplectic algebra note that the 
spin-orbit decomposition (\ref{eq:SpinOrbitE}) allows for an easy 
identification  of  Casimir operators of Gel'fand type \cite{Low2002}. 
Define
\begin{align}
\label{eq:CasimirGeneric}
{(e^{(n+1)})^\mu}_\nu = & \, {(e^{(n)})^\mu}_\rho  {e^\rho}_\nu , \quad 
{(e^{( 1)})^\mu}_\nu \equiv {e^\mu}_\nu, \nonumber \\
\mbox{then} \quad  C_{n} =  & \, \mbox{tr}(e^{(n)}) =  {(e^{(n)})^\mu}_\mu;
\end{align}
tensorially (from (\ref{eq:EeCommutators}a) these traces are $U(3,1)$ 
invariants; however from
(\ref{eq:EeCommutators}c) they are trivially also quaplectic Casimirs. 
Explicitly, we have for example with $U = {E^\mu}_\mu$
(compare (\ref{eq:SpinOrbitE}))
\begin{align}
\label{eq:QuadraticCasimirAll}
 C_1 = &\, U I - \, \textstyle{\frac 12} \big({P^\mu P_\mu} +  {X^\mu X_\mu}
\big) .
\end{align}

The quaplectic algebra itself can be re-written in a tensor form which 
identifies its Lorentz (and Poincar\'{e}) subalgebras. Identification of the Poincar\'{e} subalgebra depends on the choice of abelian four-vector operator, which can be either $X^\mu$ or $P_\nu$. The generators of the Lorentz group are then $L_{\mu\nu}=i(E_{\mu\nu}-E_{\nu\mu})$, and the remaining generators (of reciprocal boost transformations) form a symmetric tensor, $M_{\mu\nu}=E_{\mu\nu}+E_{\nu\mu}$, and the commutation relations amongst the homogeneous generators read
\begin{align}
\label{eq:ExtendedQplecticCR}
{[} L_{\kappa \lambda}, L_{\mu \nu} {]} = & \, i\left( \eta_{\lambda 
\mu}  L_{\kappa \nu} - \eta_{\kappa \mu} L_{\lambda \mu}  - 
\eta_{\lambda \nu}  L_{\kappa \mu} + \eta_{\kappa \nu}  L_{\lambda \mu} 
\right),  \nonumber \\
{[} L_{\kappa \lambda}, M_{\mu \nu} {]} = & \,i\left( \eta_{\lambda 
\mu}  M_{\kappa \nu} - \eta_{\kappa \mu} M_{\lambda \mu}  + 
\eta_{\lambda \nu}  M_{\kappa \mu} - \eta_{\kappa \nu}  M_{\lambda \mu} 
\right),  \nonumber \\
{[} M_{\kappa \lambda}, M_{\mu \nu} {]} = & \, \left( \eta_{\lambda 
\mu}  M_{\kappa \nu} + \eta_{\kappa \mu} M_{\lambda \mu}  + 
\eta_{\lambda \nu}  M_{\kappa \mu} + \eta_{\kappa \nu}  M_{\lambda \mu} 
\right), 
\end{align}
together with the usual relations expressing the transformation laws expressing the four-vector nature of  $X^\mu$ and $P_\nu$, for example,
\begin{align}
{[} L_{\kappa \lambda}, P_\mu{]} = & \, i\left(\eta_{\lambda \mu} 
P_\kappa - \eta_{\kappa \mu} P_\lambda\right). \nonumber
\end{align}

As mentioned in \S \ref{sec:Introduction}, in the worldline model there are \emph{two} independent Heisenberg algebras in the construction. For the conserved Noether charges,
it is natural to take the usual Schr\"{o}dinger representation ${\mathcal P}_\mu \rightarrow -i \hbar \partial 
/\partial x^\mu$ acting on suitable wavefunctions on Minkowski space. Such a representation is of course equivalent to one in which the complex combinations (\ref{eq:Quadratures}) are regarded as mode operators for a countable number basis, and it is this representation which is used for the auxiliary Heisenberg algebra carrying the spin degrees of freedom in our model. From (\ref{eq:HeisenbergZ}), ${\mathbb Z}_0$ is 
to be identified with a `creation' operator, whereas each ${\mathbb Z}_i$ is identified with 
an `annihilation' operator (reflecting the sign change in the 
commutation relations of ${\mathbb Z}^\mu$ and ${\overline{{\mathbb Z}}}\mbox{}^\nu$ between 
the temporal and spatial parts):
\begin{align}
\label{eq:ZasOsc}
\left(
\begin{array}{cc}
{\mathbb Z}_0 \\
{\mathbb Z}_1 \\
{\mathbb Z}_2 \\
{\mathbb Z}_3 \\
\end{array}
\right)
=
\left(
\begin{array}{cc}
a^\dagger \\
b_1 \\
b_2 \\
b_3 \\
\end{array}
\right), \quad
\left(
\begin{array}{cc}
\overline{{\mathbb Z}}_0 \\
\overline{{\mathbb Z}}_1 \\
\overline{{\mathbb Z}}_2 \\
\overline{{\mathbb Z}}_3 \\
\end{array}
\right)
=
\left(
\begin{array}{cc}
a \\
b_1^\dagger \\
b_2^\dagger \\
b_3^\dagger \\
\end{array}
\right).
\end{align}
The general state in the number basis is then \cite{Dirac1945,Kim1991}
\begin{align}
\label{eq:ZeroOccupancy}
| {n_0,n_{1},n_{2},n_3}\rangle = & \, \frac{{a^\dagger}^{n_0}{b_1^\dagger}^{n_{1}}{b_2^\dagger}^{n_{2}}{b_3^\dagger}^{n_3}}{\sqrt{n_0!n_{1}!n_{2}!n_3!}} |{0,0,0,0} \rangle.
\end{align}

\subsection{Unitary irreducible representations of $Sp(2,{\mathbb R})$ and products of discrete series representations}
\label{subsec:Sp2Unirreps}
The generators of $Sp(2, {\mathbb R}) \cong SU(1,1) \cong SO(2,1)$ are \cite{Bargmann1947,HolmanBiedenharn1966,Wang1970,Wybourne1974} $K_\pm$, and $K_0$, with  nonzero commutation relations
\begin{align}
[K_0, K_\pm] = & \, \pm K_\pm, \quad [K_+,K_-] = -2K_0, \nonumber
\end{align}
with $K_+^\dagger = K_-$ and $K_0$ hermitean  in unitary representations. The eigenvalues of $K_0$, $k$ say, are for unitary irreducible representations of $SO(2,1)$, either integer (or half-integer for the spinor case); however for general projective representations of $SU(1,1)$, $k=E_0 + m$, for integer $m$ and real $0 \le E_0 <1$. Depending on whether the spectrum of $k$ is unbounded above or unbounded below, or unrestricted, we have the positive or negative discrete, or continuous, series representations.
The quadratic Casimir 
\begin{align}
\label{eq:Sp2Casimir}
C(Sp(2)) := & \, K_0^2 - \textstyle{\frac 12} \big( K_+K_- + K_-K_+ \big) \equiv 
K_0^2 - K_0 -  K_+K_- . 
\end{align}
takes the eigenvalue $j(j+1)$, where 
\begin{align}
k = & \, \left\{ \begin{array}{rll} -j, -j+1,-j+2, \cdots & \mbox{for}\,D_j^+; & -j = 0, \frac 12, 1, \frac 32, \cdots; \\
                       j,j-1,j-2, \cdots,& \mbox{for} \, D_j^-; & -j = 0, \frac 12, 1, \frac 32, \cdots;
                       \\
                      0, \pm 1, \pm 2, \cdots, & \mbox{for} \, D(j); & j = -\frac{1}{2} +is ,\, 0\le s < \infty .
                                             \end{array} \right.  \nonumber
\end{align}
Thus for the positive and negative discrete series representations, denoted $D^\pm_j$, $j$ is the minimal (maximal) eigenvalue of $k$, and $j<0$. The continuous (principal) series representations (with Casimir $-\frac 14 -s^2$) are denoted $D(-\frac 12 +is)$; we shall not be concerned with the additional, so-called supplementary, continuous series. Finally, the one-dimensional, trivial representation is denoted ${[}0{]}$.

Of great importance are the rules for the decomposition of tensor products of discrete series representations, namely the Clebsch-Gordan series \cite{HolmanBiedenharn1966,Wang1970,Wybourne1974}
\begin{align}
\label{eq:SameTypeCGSeries}
D^\pm_{j_1} {\otimes} D^\pm_{j_2} \rightarrow & \, \sum_{j=j_1+j_2}^\infty \hspace*{-.6cm} \raisebox{0pt}{$\oplus$} \, D^\pm_j ,
\end{align}
 for the case of two same-type discrete series, while for the tensor product of positive and negative discrete series representations $D^+_{j_1}$  and $D^-_{j_2}$ we have 
\begin{align}
\label{eq:DiffTypeCGSeries}
\mbox{(for $|j_1|>|j_2|$)} \qquad  
D_{j_1}^+ \otimes D_{j_2}^-  \rightarrow & \,  \left( D_{j_1-j_2}^+ \oplus D_{j_1-j_2 +1}^+ \oplus \cdots\right) \oplus \int ds D(-\textstyle{\frac 12}+is), \nonumber \\
\mbox{(for $|j_2|>|j_1|$)} \qquad  
D_{j_1}^+ \otimes D_{j_2}^-  \rightarrow & \,  \left( D_{j_2-j_1}^- \oplus D_{j_2-j_1 +1}^- \oplus \cdots \right)  \oplus \int ds D(-\textstyle{\frac 12}+is),
\end{align}
where the last, $m\!+\!1$-th, term of the set of discrete series is $D^+_{ j_1- j_2+m}$,
or $D^-_{ j_2- j_1+m}$,
such that $-1 \le \pm j_1\mp j_2+m <0$, respectively; 
for $SO(2,1)$, this is just $-1$ or $-\frac 12$. For the degenerate case $j_1 = j_2=j$, it turns out that $D_{j}^+ \otimes D_{j}^-$ reduces as the integral over the continuous series only, \emph{together with} a copy of the one-dimensional, trivial representation.

The representation structure is easily seen for the case of a single oscillator space, with modes $b$ and $b^\dagger$ say. Defining as usual $b|0\rangle=0$, and with $[b,b^\dagger]=1$, it is well known \cite{HolmanBiedenharn1966,Wybourne1974,Gerry2004} that the Fock space decomposes with respect to the $Sp(2,{\mathbb R})$ algebra generated by 
\begin{align}
\label{eq:SimpleSp2}
K_+ = & \, \textstyle{\frac 12} (b^\dagger){}^2, \quad K_- = \textstyle{\frac 12} b^2, \quad K_0 = \textstyle{\frac 12} b^\dagger b + \textstyle{\frac 14} , 
\end{align}
into the direct sum of two (projective) unirreps of the \emph{positive} discrete series $D^+_{- 1/4} \oplus D^+_{- 3/4}$. Note that the Casimir eigenvalue is $j(j+1) = -3/16$ for both cases.
On the other hand for the \emph{opposite} identifications, for an equivalent oscillator space with modes $a, a^\dagger$, say, we can take instead
\begin{align}
\label{eq:OppositeSp2}
K_- = & \, \textstyle{\frac 12} (a^\dagger){}^2, \quad K_+ = \textstyle{\frac 12} a^2, \quad K_0 = -\textstyle{\frac 12} a^\dagger a - \textstyle{\frac 14}, 
\end{align}
and the Fock space is now the sum $D^-_{-1/4} \oplus D^-_{-3/4}$ of two (projective) unirreps of the \emph{negative} discrete series, again with Casimir invariant $j(j+1)=-3/16$. 

As an example of the application of the rules for tensor product decompositions, and as a simple illustration of the explicit elementary methods deployed below, let us take $\big( D^+_{-1/4} \oplus D^+_{-3/4}\big)\otimes\big( D^+_{-1/4} \oplus D^+_{-3/4}\big)$ corresponding for example to the case of two independent oscillator spaces. We have
\begin{align}
\big( D^+_{-1/4}\!\otimes\! D^+_{-1/4}\big)  \oplus \big(D^+_{-1/4}\! \otimes\! D^+_{-3/4}\big) 
\oplus & \, \big(D^+_{-3/4}\!\otimes\! D^+_{-1/4}\big)  \oplus \big( D^+_{-3/4}\! \otimes\! D^+_{-3/4}\big)
\rightarrow & \, \nonumber \\
D^+_{-1/2} \oplus &\, 2   \big( D^+_{-1}  \oplus D^+_{- 3/2}  \oplus D^+_{-2 } \oplus D^+_{-5/2 }  \oplus \cdots \big).
\label{eq:2Sp2Product}
\end{align} 

\subsection{$Sp(2,{\mathbb R})\! \times\! SO(d\!-\!1,1)$ or $Sp(2,{\mathbb R})\! \times\!SO(d)$ duality and Casimir relations}
\label{subsec:Sp2dDuals}
For the orbit classes corresponding to orthogonal group little algebras, the discovery of the correct reductions of physical states for each orbit class into unitary irreducible representations of the little algebras (with multiplicity controlled by the dual, commutant algebras), and hence the correct particle content, can be done by elementary means if the group branching rules are tracked instead, via the decompositions with respect to appropriate combinations of the four independent oscillator algebras (\ref{eq:ZasOsc}) (for each direction in Minkowski space).  A crucial property of these dual algebras and their little algebra partners (see below) is that the respective Casimir invariants are \emph{identical} (up to factors), a fact which greatly facilitates the recognition of unirreps.

The general definition of the dual $Sp(2,{\mathbb R})$ commuting algebras in the chains $Sp(2d,{\mathbb R}) \supset  Sp(2,{\mathbb R})\! \times\! SO(d\!-\!1,1)$ or 
$Sp(2d,{\mathbb R}) \supset Sp(2,{\mathbb R})\! \times\!SO(d)$, for $d = 3$ (for $SO(3)$ or $SO(2,1)$),  $d=4$ (for $SO(3,1)$), and also $d=2$ for intermediate calculations, is as follows. Consider the Heisenberg algebra generators ${\mathbb Z}^a$, $\overline{{\mathbb Z}}_b$ for $a,b \in I \subseteq \{0,1,2,3\}$ corresponding to the subset of directions in Minkowski space appropriate to the little algebra construction for standard $4$-momentum $\!\pcirc\!$, with $d=|I|$. Thus the little algebra is spanned by the orthogonal generators ${\mathbb L}_{ab}$ corresponding to $SO(d-r,r)$ with $r=0$ or $1$ depending on $I$. Define the \emph{contracted} combinations
\begin{align}
\label{eq:KfromZ}
K_+ = & \, \textstyle{\frac 12} \overline{\mathbb Z}{}^a \overline{\mathbb Z}_a, \quad 
K_- = \textstyle{\frac 12} {\mathbb Z}{}^a {\mathbb Z}_a, \quad 
K_0 =  - \textstyle{\frac 12} \overline{\mathbb Z}{}^a  {\mathbb Z}_a + \textstyle{\frac 14}{d}. 
\end{align}
These generators provide the desired $Sp^{(I)}(2,{\mathbb R})$ Lie algebra and obviously commute with the ${\mathbb L}_{ab}$; these algebras are both subalgebras of $Sp(2d,{\mathbb R})$ which is generated by all quadratic combinations of the ${\mathbb Z}^a$, $\overline{{\mathbb Z}}_b$ (see for example \cite{JarvisMorgan2006}). Moreover, it is easily shown that the Casimir  (\ref{eq:Sp2Casimir}), is related to the orthogonal group Casimir 
$C(SO(d-r,r)) := \textstyle{\frac 12}{\mathbb L}^{ab} {\mathbb L}_{ab}$ by
\begin{align}
\label{eq:DualCasimirs}
4 C(Sp^{(I)}(2,{\mathbb R})) = & \,  C(SO(d-r,r)) + \textstyle{\frac{1}{4}}{d(d-4)}
\end{align}
(\emph{independently} of the signature of the orthogonal metric).


As an illustration of the method to be elaborated in \S \S \ref{subsec:Massive}, \ref{subsec:Spacelike}, and \ref{subsec:Null} below, consider the implications of (\ref{eq:DualCasimirs}) for the case of $d=2$ in (\ref{eq:2Sp2Product}) above. For $r=0$ and Euclidean signature, we have the group branchings
\begin{align}
\label{eq:Sp4Euclidean}
Sp(4, {\mathbb R}) \supset & \, SO(2) \times Sp(2,{\mathbb R}); \nonumber \\
\mbox{and} \qquad 
Sp(4, {\mathbb R}) \supset & \, Sp(2,{\mathbb R}) \times Sp(2,{\mathbb R})
				\supset  Sp(2,{\mathbb R}), 
				\end{align}
whereas in the case $r=1$, and Minkowski signature we have 
\begin{align}
\label{eq:Sp4Minkowski}
Sp(4, {\mathbb R}) \supset & \, SO(1,1) \times Sp(2,{\mathbb R}); \nonumber \\
\mbox{and} \qquad 
Sp(4, {\mathbb R}) \supset & \, Sp(2,{\mathbb R}) \times Sp(2,{\mathbb R})
				\supset  Sp(2,{\mathbb R}). 
				\end{align}
For each signature type, the final $Sp(2,{\mathbb R})$ group is the same in both branchings; in the second alternative, however, the result of the reduction to the final $Sp(2,{\mathbb R})$ group is precisely (\ref{eq:2Sp2Product}). Thus the dimensions and degeneracies of the occurring $SO(2)$ or $SO(2,1)$ unirreps are manifested directly in the multiplicities displayed in (\ref{eq:2Sp2Product}). 

In the $SO(2)$ case, taking for example $a,b = 1,2$, the generator is simply $iL_{12}$, with 
one dimensional unirreps ${[}m{]}$ (the contragredient being ${[}\!-\!m{]}$), and Casimir $L_{12}^2 = m^2$. The double multiplicity of $Sp(2,{\mathbb R})$ discrete series thus signals the repetition of a given unirrep, or rather the occurrence of it together with its contragredient. Using (\ref{eq:DualCasimirs}) for $d=2$,
\begin{align}
m^2 = & \, C(SO(2)) = 4C(Sp(2,{\mathbb R})) +1, \nonumber
\end{align}
we finally recover the desired $SO(2) \times Sp(2,{\mathbb R})$ tensor product decomposition in the form
\begin{align}
\big( D^+_{-1/4} \oplus D^+_{-3/4}\big)\otimes\big( D^+_{-1/4} \oplus D^+_{-3/4}\big) \rightarrow & \,
\big({[}0{]}\!\otimes \!D^+_{-1/2} \big)\oplus \left(
\sum_{\ell = 2}^\infty \hspace*{-.1cm} \raisebox{0pt}{$\oplus$} 
\big({[}\ell\!-\!1{]}\! \otimes \! D^+_{-\ell/2}\big) \oplus \big({[}\!-\!(\ell\!-\!1){]}\! \otimes \! D^+_{-\ell/2}  \big) \right).\nonumber \\
\label{eq:2OscillatorEuclidean}
\end{align}

\subsection{Massive states $p\!\cdot\!p >0$}
\label{subsec:Massive}
For $\!\pcirc = mc(1,0,0,0)$ the little algebra $SO(3)$ generators ${\mathbb L}_{ij}$, $i,j = 1,2,3$
commute with $Sp^{(123)}(2,{\mathbb R})$ with branching rules extending those of 
(\ref{eq:Sp4Euclidean}), namely
\begin{align}
\label{eq:Sp6Euclidean}
Sp(6, {\mathbb R}) \supset & \, SO(3) \times Sp^{(123)}(2,{\mathbb R}); \nonumber \\
\mbox{and} \qquad 
Sp(6, {\mathbb R}) \supset & \, Sp^{(1)}(2,{\mathbb R}) \times Sp^{(2)}(2,{\mathbb R})\times Sp^{(3)}(2,{\mathbb R}) \supset  Sp^{(12)}(2,{\mathbb R})\times Sp^{(3)}(2,{\mathbb R})
\nonumber \\
\supset & \,  Sp^{(123)}(2,{\mathbb R}) 
\end{align}
so that the method leading to (\ref{eq:2Sp2Product}), (\ref{eq:2OscillatorEuclidean}) must be extended from the reduction of the two-oscillator tensor product space, to the reduction after carrying out the tensor product with a third oscillator space, and using the relation between Casimir invariants to identify the constituents with respect to $SO(3)$.

The results are as follows. By direct application of (\ref{eq:SameTypeCGSeries}), we have from
(\ref{eq:2Sp2Product}),
\begin{align}
\big( D^+_{-1/4} \oplus D^+_{-3/4}\big)^2\otimes\big( D^+_{-1/4} \oplus D^+_{-3/4}\big)
\cong & \, \nonumber \\
D^+_{-3/4} \oplus &\, 3  D^+_{-5/4}  \oplus 5 D^+_{- 7/4}  \oplus 7D^+_{-9/4 } \oplus 9 D^+_{-11/4 }  \oplus \cdots ,
\label{eq:3Sp2Product}
\end{align} 
and using (\ref{eq:DualCasimirs}) for $d=3$, with the $SO(3)$ Casimir invariant for unitary irreducible representation ${[}\ell{]}$ (of dimension $2\ell+1$) being in the form
\begin{align}
\ell(\ell+1) = & \, C(SO(3)) = 4C(Sp^{(123)}(2,{\mathbb R})) +\textstyle{\frac 34}, \nonumber
\end{align}
the final reduction to $SO(3)\times Sp^{(123)}(2,{\mathbb R})$ reads
\begin{align}
\label{eq:3OscillatorEuclidean}
\big( D^+_{-1/4} \oplus D^+_{-3/4}\big)^3 \rightarrow & \,
\big({[}0{]}\!\otimes \!D^+_{-3/4} \big) 
\oplus\big({[} 1{]} \!\otimes \!D^+_{-5/4} \big)\oplus\big({[} 2{]} \!\otimes \!D^+_{-7/4} \big)\oplus\big({[} 3{]} \!\otimes \!D^+_{-9/4} \big)\oplus \cdots
\nonumber \\
\cong & \,
\sum_{\ell = 0}^\infty \hspace*{-.1cm} \raisebox{0pt}{$\oplus$} 
\big({[} \ell{]} \! \otimes \! D^+_{-\!\ell/2\!-\!3/4}\big) .
\end{align}
\subsection{Tachyonic states $p\!\cdot\!p <0$}
\label{subsec:Spacelike}
For $\!\pcirc = \!\pcirc\!\!\!\!{}_z(0,0,0,1)$ the little algebra $SO(2,1)$ generators ${\mathbb L}_{ab}$, $a,b = 0,1,2$
commute with $Sp^{(012)}(2,{\mathbb R})$ with branching rules extending those of 
(\ref{eq:Sp4Euclidean}), namely
\begin{align}
\label{eq:Sp6Minkowski}
Sp(6, {\mathbb R}) \supset & \, SO(2,1) \times Sp^{(012)}(2,{\mathbb R}); \nonumber \\
\mbox{and} \qquad 
Sp(6, {\mathbb R}) \supset & \, Sp^{(0)}(2,{\mathbb R}) \times Sp^{(12)}(4,{\mathbb R})
                                 \supset  Sp^{(0)}(2,{\mathbb R})\times Sp^{(12)}(2,{\mathbb R})\times SO(2)
                                \nonumber \\                                
\supset & \,  SO(2) \times Sp^{(012)}(2,{\mathbb R}).
\end{align}
In this case the final two-oscillator space decomposition (\ref{eq:2OscillatorEuclidean}) must again be extended by carrying out the tensor product with a third oscillator space. However, in contrast to the $SO(3)$ decomposition, in this case from (\ref{eq:ZasOsc}), (\ref{eq:KfromZ}), the realisation of the $Sp^{(0)}(2,{\mathbb R})$ algebra requires the identification of the oscillator Fock space with the direct sum 
$D^-_{-1/4} \oplus D^-_{-3/4}$ of \emph{negative} discrete series representations, and the rules (\ref{eq:DiffTypeCGSeries}) must be used. The unirreps of $SO(2)$ occurring can however be used along with knowledge of the spectrum of the diagonal generator $i{\mathbb L}_{12}$ in unirreps of $SO(2,1)\cong SU(1,1)$, and the relation between Casimir invariants, again
\begin{align}
 C(SO(2,1)) = & \, 4C(Sp^{(012)}(2,{\mathbb R})) +\textstyle{\frac 34}, \nonumber
\end{align}
to identify the constituents with respect to $SO(2,1)$.
The result for the $SO(2)\times Sp^{(012)}(2,{\mathbb R})$ decomposition is 
\begin{align}
\big( D^+_{-1/4} \oplus D^+_{-3/4}\big)^2 \otimes 
 \big( D^-_{-1/4} \oplus D^-_{-3/4}\big) \rightarrow & \, 
\big({[}0{]}\otimes(D^+_{-1/4} \oplus D^-_{-1/4})\big) \oplus \nonumber \\
\oplus \big(({[}\!+\!1{]}\oplus{[}\!+\!2{]}\oplus{[}\!+\!3{]} \oplus \cdots)\otimes D^+_{-1/4} \big)
\oplus & \, \big(({[}\!-\!1{]}\oplus{[}\!-\!2{]}\oplus{[}\!-\!3{]} \oplus \cdots)\otimes D^+_{-1/4} \big)  \nonumber \\
\oplus \big(({[}\!+\!2{]}\oplus{[}\!+\!3{]}\oplus{[}\!+\!4{]} \oplus \cdots)\otimes D^+_{-3/4} \big)
\oplus & \, \big(({[}\!-\!4{]}\oplus{[}\!-\!3{]}\oplus{[}\!-\!4{]} \oplus \cdots)\otimes D^+_{-3/4} \big)  \nonumber \\
\oplus 2\big((\cdots {[}\!-\!2{]}\oplus{[}\!-\!1{]}\oplus{[}0{]}\oplus{[}\!+\!1{]}&\oplus{[}\!+\!2{]} \oplus \cdots)\otimes
\int D(-\textstyle{\frac 12}+is') ds' .
\label{eq:3OscillatorMinkowski}
\end{align}
from which we finally recover the $SO(2,1)\times Sp^{(012)}(2,{\mathbb R})$ decomposition
\begin{align}
\big( D^+_{-1/4} \oplus D^+_{-3/4}\big)^2 \otimes 
 \big( D^-_{-1/4} \oplus D^-_{-3/4}\big) \rightarrow & \, 
\nonumber \\
\big({[}0{]}\otimes(D^+_{-\frac 14} \oplus D^-_{-\frac 14})\big) \oplus   \left( \sum_{\ell = 1}^\infty
\big(D^+_{-\!\ell}\! \otimes \! D^+_{-\!\frac 12\ell\!+\!\frac 14}\big)\oplus\right. & \,\left. \big(D^-_{-\!\ell}\! \otimes \! D^+_{-\!\frac 12\ell\!+\!\frac 14}\big) \right) \nonumber \\
\oplus  \, 2\big( \int D(-\textstyle{\frac 12}+is) \otimes D(-\textstyle{\frac 12}+2is) ds \big). &
\label{eq:3OscillatorMinkowskiFinal}
\end{align}
\subsection{Null states $p \equiv 0$}
\label{subsec:Null}
The relevant branching chain is
\begin{align}
Sp(8,{\mathbb R}) \supset & \, {Sp}^{(0123)}(2,{\mathbb R})\times  SO(3,1)
\end{align}
so that the degeneracy of null `particle' states is controlled by the decomposition with respect to ${Sp}^{(0123)}(2,{\mathbb R})$. The structure of this Fock space (comprising the $a$ mode together with all three of the $b$ modes) with respect to the total $Sp^{(0123)}(2, {\mathbb R})$ is thus that of the reduction of the tensor product decomposition ${Sp}^{(0)}(2) \times SO(3) \times Sp^{(123)}(2)$, 
\begin{align}
& \, (D^-_{-\frac 14} \! \oplus \! D^-_{-\frac 34}) \! \otimes \! \left( \sum_{\ell = 0}^\infty \hspace*{-.2cm} \oplus\,  [\ell]\!\otimes\!D^+_{-\ell - \frac 34} \right), 
\end{align}
resulting in an $SO(3) \times Sp^{(0123)}(2)$ decomposition from which can be inferred the 
$SO(3,1) \times Sp^{(0123)}(2)$ branching, knowing the structure of unirreps of $SO(3,1)$ in an $SO(3)$ basis. 
At the $SO(3) \times Sp^{(0123)}(2)$ level this reads
\begin{align}
\sum_{\ell = 0}^\infty \hspace*{-.2cm} \oplus\,  [\ell]\!\otimes\! \left( D^+_{-\frac 12 \ell- \frac 12}\!\oplus\! D^+_{-\frac 12 \ell+ \frac 12} \!\oplus\! \cdots
\!\oplus\! D^+_{-\frac 12} \! \oplus2  \! \int ds D(-\textstyle{\frac 12} + is) \right)
\end{align}
which can be rearranged as
\begin{align}
 2 \! \left( \! \int ds D(-\textstyle{\frac 12} + is) \otimes  \displaystyle{\sum_{\ell = 0}^\infty{}} \hspace*{-.3cm}\!\oplus [\ell] \right)&\, \oplus
 \left( D^+_{-\frac 12 \ell - \frac 12} \!\otimes\! \sum_{\ell' = \ell}^\infty \hspace*{-.2cm}\oplus [\ell ']\right).
\end{align}
Unirreps of $SO(3,1)$ are denoted $D[k_0,c]$ according to the parametrisation of the eigenvalue
$k_0^2+c^2-1$ of the aforementioned $C(SO(3,1))$ Casimir for this case, together with the eigenvalue $ik_0c$ of the second quadratic Casimir $C'(SO(3,1) = \frac 18 \epsilon^{\mu\nu\rho\sigma}L_{\mu\nu}L_{\rho\sigma}$ (see \cite{Gelfand1963}). In either case $2|k_0|$ is integral, and the angular momentum ($SO(3)$) spectrum is $\ell = |k_0|$, $|k_0|+1, \cdots $. For Lorentz transformations associated with orbital angular momentum (as is the case for ${\mathbb L}_{\mu\nu}$ in the auxiliary space) $C'$  vanishes, and the relevant representations are in the principal series with either $k_0=0$, namely $D[0,-2is]$, or $c=0$, namely $D[\ell,0]$; the $C$ eigenvalues being $-1-4s^2$ and $\ell^2-1$, respectively. Given the interrelationship (\ref{eq:DualCasimirs}) for this case, 
\begin{align}
C(SO(3,1)) = & \, 4C(Sp^{(0123)}(2,{\mathbb R})), \nonumber
\end{align}
it is easily checked therefore that the reduction with respect to $Sp^{\equiv 0}(2)\times SO(3,1)$ is simply
\begin{align}
 2 \! \left( \! \int ds D(-\textstyle{\frac 12} + is) \otimes  D[0,-2is] \right)&\, \oplus
\left(  \sum_{\ell = 0}^\infty \hspace*{-.2cm}\oplus  D^+_{-\frac 12 \ell - \frac 12} \!\otimes\! D[\ell,0]\right ).
\end{align}
\subsection{${\mathcal E}(2) \times {\mathbb E}(2)$ duality and massless states $p\!\cdot\!p =0$}
\label{subsec:E2Cases}
In the massless case the little algebra ${\mathcal L}^{(=0)} ={\mathcal E}(2)$ involves noncompact generators such as ${\mathbb L}_{01}+{\mathbb L}_{03}$ whose diagonalisation requires some analytical technicalities \cite{LindbladNagel1970}. This algebra does not admit an $Sp(2,{\mathbb R})$ type commutant, but instead has a dual ${\mathbb E}(2)$ of the same Euclidean type either within $Q(3,1)$ or $Sp(8,{\mathbb R})$.

In a standard basis the Lie algebra $E(2)$ has generators $T_\pm$ and $M$ (the compact $SO(2)$ generator), with commutation relations
\begin{align}
{[}M,T_\pm{]} = & \, \pm T_\pm, \qquad {[}T_+, T_-{]} =0,
\end{align}
and Casimir invariant $C(E(2)) = T_+T_- = T_1^2 + T_2^2$ where $T_\pm = T_1 \pm iT_2$ in terms of hermitean translation generators in Cartesian coordinates. For continuous series unirreps $D{(}\pi{)}$, The Casimir eigenvalue is $\pi^2$, $0<\pi<\infty$, and with respect to a standard basis with  $M$ diagonal (and integer, or half-odd integer for spin representations), matrix elements
\begin{align}
T_\pm = & \, \pi |\pi,m\pm 1 \rangle, \quad M |\pi,m\rangle = m |\pi,m\rangle, \nonumber \\
\mbox{and} \qquad \langle \pi',m' | \pi,m\rangle = & \, \delta(\pi'-\pi)\delta_{m'm}.
\end{align}
On the other hand the discrete series representations ${[}m{]}$ are one dimensional helicity states 
$M |m\rangle = m |m\rangle$ and vanishing Casimir.

A useful step to understanding the full decomposition into massless states in the present, quaplectic  realisation, is to consider the case of a two-dimensional oscillator system (see (\ref{eq:ZasOsc}) for index set $I = {\{}1,2{\}}$). For this system $M = iL_{12}$ as usual, and the translations $T_\pm$ are nothing but the complex momentum combinations $P_\pm = P_1\pm iP_2$. Adopting the appropriate raising and lowering combinations for circular oscillators,
$b_\pm = \frac{1}{\sqrt{2}}(b_1 \mp b_2)$ we have
\begin{align}
M = & \, b_+^\dagger b_+ - b_-^\dagger b_-, \quad 
P_\pm = b_\pm - b_\mp^\dagger, \nonumber, \\
P_+P_- = & \, 1 + b_+^\dagger b_+ + b_-^\dagger b_- - b_+^\dagger b_-^\dagger - b_+ b_- .
\label{eq:E2Euclidean}
\end{align}
In spherical polar coordinates $(r,\phi)$, we have $M \rightarrow -i\partial/\partial \phi$,
and $P_\pm \rightarrow -ie^{\pm i\phi}(\partial /\partial r \pm (i/r)\partial/\partial \phi)$, and diagonalisation leads to the well known eigenfunctions $\langle r,\phi|\pi,m\rangle \cong J_m(\pi r) \exp(im\phi)$ so that the oscillator space decomposes into a direct integral over $0<\pi<\infty$ of $D(\pi)$.

In the present quaplectic realisation, the algebraic structure is best seen via a complex tetrad basis relative to the reference lightlike four-momentum, namely
\begin{align}
\stackrel{\circ}{\raisebox{0pt}{\scalebox{1.05}{$p$}}}= (1,0,0,1), \quad & \, 
\stackrel{\circ}{\raisebox{0pt}{\scalebox{1.05}{$p$}}}{}'= (1,0,0,-1), \quad 
u^{+} = \textstyle{\frac{1}{\sqrt{2}}}(0,1,  i, 0)= \overline{u}{}^-. \nonumber
\end{align}
Then defining for four-vectors $v,w,\cdots$ ${\mathbb Z}(v) = {\mathbb Z}^\mu v_\mu$, ${\mathbb L}(v,w) = v^\mu w^\nu {\mathbb L}_{\mu\nu}$ $= - {\mathbb L}(w,v)$, and so on, we have the 
generators:
\begin{align}
\label{eq:E2algebras}
{\mathcal E}(2):\qquad T_\pm =& \,  {\mathbb L}(\stackrel{\circ}{\raisebox{0pt}{\scalebox{1.05}{$p$}}},{u}^\pm), \quad {M} = 
{\mathbb L}({u}^+,u^-); \nonumber \\
{\mathbb E}(2): \qquad {\mathbb T}_+ = & \, \overline{\mathbb Z}(\stackrel{\circ}{\raisebox{0pt}{\scalebox{1.05}{$p$}}}),
\quad {\mathbb T}_- = {\mathbb Z}(\stackrel{\circ}{\raisebox{0pt}{\scalebox{1.05}{$p$}}}), \quad
{\mathbb M} = \overline{\mathbb Z}\cdot {\mathbb Z}; 
\end{align}
note that an alternative commutant can indeed be found within the full $Sp(8,{\mathbb R})$ symplectic algebra, by defining simply
${\mathbb T}'_\pm={\mathbb T}_\pm^2$, ${\mathbb M}' = 2{\mathbb M}$.

Explicit forms of the generators can be constructed by referring to (\ref{eq:ZasOsc}); it is convenient to re-label $b_3 \equiv c$, $b^\dagger_3 \equiv c^\dagger$ along with $a$, $a^\dagger$ in order to distinguish the longitudinal $0,3$ (`$\|$') and transverse $1,2$ or $+,-$ directions.
We have
\begin{align}
{\mathbb T}_+ =& \, a^\dagger -c,\quad {\mathbb T}_-  = a - c^\dagger , \nonumber \\
{\mathbb M} = & \, (a^\dagger a - c^\dagger c) - (b_+^\dagger b_+ + b_-^\dagger b_-) ; \nonumber \\
{\mathbb T}_+{\mathbb T}_+ = & \,1+ a^\dagger a + c^\dagger c
-a^\dagger c^\dagger-ac; \label{eq:E2AuxilAsOsc} \\
\mbox{whereas}\qquad \qquad \quad T_\pm = & \, (a^\dagger -c)b_\pm^\dagger 
-(a - c^\dagger)b_\mp \equiv {\mathbb T}_+b_\pm^\dagger -{\mathbb T}_- b_\mp ,\nonumber \\
M = & \, b_+^\dagger b_+ - b_-^\dagger b_-,
  \nonumber \\
\mbox{and finally} \qquad \qquad T_+T_- = & \, 
\big({\mathbb T}_-^2 b_+^\dagger b_-^\dagger + {\mathbb T}_+^2b_+ b_-) -{\mathbb T}_+{\mathbb T}_-\big(1 + b_+^\dagger b_+ + b_-^\dagger b_- \big).
\label{eq:E2LittleAsOsc}
\end{align}
There is an obvious equivalence between (\ref{eq:E2Euclidean}), (\ref{eq:E2AuxilAsOsc}) which can be used to reduce the $0,3$ oscillator space to a direct integral of continuous series ${\mathbb E}(2)$ unirreps $D(\pi_\|)$.
Moreover, from (\ref{eq:E2LittleAsOsc}) on such states $|\psi; \pi_\|,m_\|\rangle \equiv |\psi\rangle \otimes | \pi_\|,m_\|\rangle$, where $|\psi\rangle$ is an arbitrary state of the $1,2$ oscillator space, we have evidently
\begin{align}
T_\pm|\psi; \pi_\|,m_\|\rangle = & \, 
\pi_\| \big( b_\pm^\dagger -b_\mp \big)  |\psi;  \pi_\|,m_\| \rangle ; \nonumber \\
M|\psi; \pi_\|,m_\|\rangle = & \, \big(b_+^\dagger b_+ - b_-^\dagger b_-\big) |\psi; \pi_\|,m_\|\rangle
\end{align}
so that (if necessary by rescaling generators $T_\pm \rightarrow T_\pm/ \pi_\|$), essentially the \emph{same} diagonalisation renders a direct integral of states of the form
\begin{align}
|\pi \pi_\|, m;  \pi_\|,m_\| \rangle,
\end{align}
with transverse quantum number (the helicity, the difference between the positive and negative circular oscillator mode numbers) $m$, and auxiliary mode number $m_\|$.
The final projection onto physical states obeying the constraint is thus trivial in the present case: we require simply $\Delta = m_\|$. 
\end{appendix}
\section*{}
\begin{table}[tbp]
  \centering 
  \begin{tabular}{lllll}
  \hline
  \raisebox{-.3cm}{\rule{0cm}{.8cm}} & $\pcirc $ & ${\mathcal L}$ & ${\mathbb L}$ & $\frac 12(\widehat{N}_1 \!+\! \widehat{N}_2 \!+\! \widehat{N}_3\! -\! \widehat{N}_0)\! +\! \frac 12$ \\
   \hline
 \raisebox{-.3cm}{\rule{0cm}{.8cm}}  $>0$ &  $(1,0,0,0)$ & $SO(3)$ & $Sp^{(0)}(2,{\mathbb R})+Sp^{(123)}(2,{\mathbb R})$ & $K_0^{(0)}+ K_0^{(123)}$ \\
 \raisebox{-.3cm}{\rule{0cm}{.8cm}}  $<0$ &  $(0,0,0,1)$ & $SO(3)$ & $Sp^{(012)}(2,{\mathbb R})+Sp^{(3)}(2,{\mathbb R})$ & $K_0^{(012)}+ K_0^{(3)}$\\
 \raisebox{-.3cm}{\rule{0cm}{.8cm}}  $\equiv 0$ &  $(0,0,0,0)$ & $SO(3,1)$ & $Sp^{(0123)}(2,{\mathbb R})$ & $K_0^{{(0123)}}$ \\
 \raisebox{-.3cm}{\rule{0cm}{.8cm}}  $= 0$ &  $(1,0,0,1)$ & ${\mathcal E}(2)$ & ${\mathbb E}(2)$ &
 ${\mathbb M}$ \\
   \hline
   \end{tabular}
\caption{Standard frame momentum coordinates, 
little algebras and duals, 
 ${\mathcal L} \times {\mathbb L}$, for $p\!\cdot\!p >0$ (massive), $<0$ (tachyonic), $p\equiv 0$ (null) and $=0$ (massless), together with the constraint operator to be projected onto
 eigenvalue $\Delta$, expressed in terms of the diagonal generators of the dual algebras (for discussion and notation see \S 3 and appendix, \S A ).}\label{tab:LittleAlgTab}
\end{table}
\begin{table}[tbp]
  \centering 
    \begin{tabular}{lllll}
  \hline
  \raisebox{-.3cm}{\rule{0cm}{1cm}}  $p{\!}\cdot{\!}p$ & ${\mathcal L}$ &  ${\mathbb L}$ &  & \\
   \hline
 \raisebox{-.3cm}{\rule{0cm}{1cm}}  $>0$ &  $SO(3)$ &  $Sp^{(123)}(2,{\mathbb R})$  &$Sp^{(0)}(2,{\mathbb R})$ &  \\
 \raisebox{-.3cm}{\rule{0cm}{1cm}}  &  ${[}\ell{]}$ &  $D^+_{-\frac 12\ell - \frac 34}$  &
 $D^-_{-\frac 14}\!\oplus\! D^-_{-\frac 34}$ & ${\sum_{\ell = 0}^\infty}$ \\
 \hline
  \raisebox{-.3cm}{\rule{0cm}{1cm}}  $<0$ &    $SO(2,1)$ &  $Sp^{(012)}(2,{\mathbb R})$ &$Sp^{(3)}(2,{\mathbb R})$  &\\
  \raisebox{-.3cm}{\rule{0cm}{1cm}}    &    ${[}0{]}$ &  $D^+_{-\frac 14}\!\oplus\! D^-_{-\frac 14}$&
  $D^+_{-\frac 14}\!\oplus\! D^+_{-\frac 34}$ & \\
 \raisebox{-.3cm}{\rule{0cm}{1cm}}   & $D^+_{\ell}$ &  $D^+_{-\frac 12\ell + \frac 14}$& 
  $D^+_{-\frac 14}\!\oplus\! D^+_{-\frac 34}$ & ${\sum_{\ell = 1}^\infty}  $ \\
 \raisebox{-.3cm}{\rule{0cm}{1cm}} & $D^-_{\ell}$ &  $D^+_{-\frac 12\ell + \frac 14}$& 
 $D^+_{-\frac 14}\!\oplus\! D^+_{-\frac 34}$ & ${\sum_{\ell = 1}^\infty}  $ \\
 \raisebox{-.3cm}{\rule{0cm}{1cm}} & $D(-\frac 12\!+\! is)$ &  $2\,D(-\frac 12\!+\! 2is)$ & 
 $D^+_{-\frac 14}\!\oplus\! D^+_{-\frac 34}$ &${\int_{0}^\infty}  ds $ \\
\hline
      \raisebox{-.3cm}{\rule{0cm}{.9cm}}  $\equiv0$ & $SO(3,1)$  & $Sp^{(0123)}(2,{\mathbb R})$ &  & \\
    \raisebox{-.3cm}{\rule{0cm}{.9cm}}   & $D{[}\ell,0{]}$  & $D^+_{-\frac 12 \ell -\frac 12}$ &  &${\sum_{\ell = 0}^\infty}  $ \\
   \raisebox{-.3cm}{\rule{0cm}{.9cm}}   & $D{[}0,-2is{]}$  & $2\,D(-\frac 12\!+\! is)$ &  &${\int_{0}^\infty}  ds $ \\
      \hline
    \raisebox{-.3cm}{\rule{0cm}{.9cm}}  $=0$ & ${\mathcal E}(2)$   & ${\mathbb E}(2)$ &   &  \\
    \raisebox{-.3cm}{\rule{0cm}{.9cm}}   & $D(\pi \pi_\|)$   & $D( \pi_\|)$ &   &${\int \! \! \! \int_{0}^\infty}  d\pi d\pi_\| $ \\ 
   \hline
   \end{tabular}
  \caption{Branches of the mass-squared spectrum of physical states, $p\!\cdot\!p >0$ (massive), $=0$ (massless), $<0$ (tachyonic) and $p\equiv 0$ (null) together with irreducible representations of the
 respective little algebras and associated degeneracy algebras, to be projected with respect to the constraint eigenvalue $\Delta$ (for discussion and notation see table \ref{tab:LittleAlgTab}, \S 3 and appendix, \S A ).  }\label{tab:Results}
\end{table}

\begin{thebibliography}{99}

\bibitem{GovaertsEtAl2007}
Jan Govaerts, Peter D.~Jarvis, Stuart O.~Morgan and Stephen G.~Low, 
\newblock {\sl World-line Quantisation of
 a Reciprocally Invariant System}, {\em J. Phys.} {\bf A40}, 12095--12111 (2007).

\bibitem{Born1949}
M.~{Born}, \newblock {\sl Elementary Particles and the Principle of Reciprocity\/},
\newblock {\em Nature} \textbf{163}, 207--208 (1949).

\bibitem{Born1949a}
M.~{Born}, \newblock {\sl Reciprocity Theory of Elementary Particles\/},
\newblock {\em Reviews of Modern Physics} \textbf{21}(3), 463--473 (1949).

\bibitem{Green1949}
H.~S.~{Green}, \newblock {\sl Quantized Field Theories and the Principle of Reciprocity\/},
\newblock {\em Nature} \textbf{163}, 208--209 (1949).

\bibitem{Green1949a}
H.~S.~Green, {\sl Theory of Reciprocity, Broken SU(3) Symmetry, and Strong Interactions\/},
Proc. Int. Conf. on Elementary Particles, Kyoto, 1965, {\it Prog. Theory. Phys.}
(1966) 159; see also \\
A.~J.~Bracken, Ph.D. Thesis, University of Adelaide (1970).

\bibitem{Low2002}
Stephen~G.~{Low}, \newblock {\sl Representations of the Canonical Group,
(the Semidirect Product of the Unitary and {W}eyl-{H}eisenberg Groups), 
Acting as a Dynamical Group on Noncommutative Extended Phase Space\/},
\newblock {\em J. Phys. A} \textbf{35}, 5711--5729 (2002).

\bibitem{Low2005a}
Stephen~G.~{Low}, \newblock {\sl Reciprocal Relativity of Noninertial Frames: Quantum
Mechanics\/}, \newblock {\em J. Phys. A} \textbf{40}, 3999--4016 (2007).

\bibitem{Low2005b}
Stephen~G.~{Low}, \newblock {\sl Reciprocal Relativity of Noninertial Frames
and the Quaplectic Group\/}, \newblock  {\em Found. Phys.} \textbf{36} (7), 1036--1069  (2006).

\bibitem{Govbook}
See for example, and references therein,\\
J.~Govaerts, {\sl Hamiltonian Quantisation and Constrained Dynamics} (Leuven University Press,
Leuven, 1991).

\bibitem{Gov2}
J.~Govaerts, {\sl The Cosmological Constant of One-Dimensional
Matter Coupled Quantum Gravity is Quantised},
Proceedings of the Third International Workshop on Contemporary
Problems in Mathematical Physics (2003), Cotonou (Republic of Benin),
eds. J. Govaerts, M. N. Hounkonnou and A. Z. Msezane (World Scientific,
Singapore, 2004), pp.~244--272 [e-print {\tt arXiv:hep-th/0408022}].

\bibitem{Wigner1939} E.~P.~Wigner , {\sl On Unitary Representations of the Inhomogeneous Lorentz Group' and elementary particle classification \/}, {\it Ann. Math. \/} {\bf 40}, 149 (1939).

\bibitem{Shamaly1974} A.~Shamaly, \newblock {\sl Invariants of Born reciprocity theory\/},  \newblock {\em J. Math. Phys. \/} {\bf 15}, 1178-1180 (1974).

\bibitem{Gov3}
See for instance, and references therein,\\
J.~Govaerts and V.~M.~Villanueva, {\sl Topology Classes of Flat $U(1)$ Bundles and
Diffeomorphic Covariant Representations of the Heisenberg Algebra \/},
{\it Int. J. Mod. Phys.\/} {\bf A15}, 4903 (2000).

\bibitem{ORaif}
L.~O'Raifeartaigh, {\sl Internal Symmetry and Lorentz Invariance\/},
{\em Phys. Rev. Lett.} {\bf 14}, 332 (1965);\\
L.~O'Raifeartaigh, {\sl Mass Differences and Lie Algebras of Finite Order\/},
{\em Phys. Rev. Lett.} {\bf 14}, 575 (1965);\\
L.~O'Raifeartaigh, {\sl Lorentz Invariance and Internal Symmetry\/},
{\em Phys. Rev.} {\bf 139}, B1052 (1965);\\
G.~Fuchs, {\sl About O'Raifeartaigh's Theorem}, {\em Annales de l'Inst. Henri Poincar\'e}
{\bf 9}(1), 7--16 (1968).

\bibitem{JarvisMorgan2006}
P.~D.~Jarvis and S.~O.~Morgan, {\sl Born Reciprocity and the Granularity
of Spacetime \/}, \newblock  {\em Found. Phys. Lett.} \textbf{19} (6), 501--517  (2006). 


\bibitem{Low2008}
Stephen~G.~{Low}, \newblock {\sl Hamiltonian relativity group for noninertial states in quantum mechanics\/}, \newblock  {\em J. Phys.}  {\bf A} (2008, to appear).

\bibitem{Wolf1975} J.~A.~Wolf,
\newblock {\sl Representations of certain semidirect product groups\/},
\newblock {\em J. Func. Anal.} \textbf{19} 339--372 (1975).

\bibitem{Dirac1945}
P.~A.~M Dirac,
\newblock {\sl Unitary representations of the {L}orentz group \/},
\newblock {\em Proc. Roy. Soc. (London)} \textbf{A183} 285--295 (1945).

\bibitem{Kim1991}
Y.~S.~ {Kim} and M.~E. {Noz},
\newblock {\sl Phase space picture of quantum mechanics: group theoretical
  approach \/}.
\newblock (World Scientific, Singapore, 1991).

\bibitem{Bargmann1947} V.~Bargmann, \newblock {\sl Irreducible unitary representations of the Lorentz group}, \newblock {\em Ann. Math.} {\bf 48}(3), 568--640 (1947).

\bibitem{HolmanBiedenharn1966} Wayne J.~Holman and Lawrence C.~Biedenharn, \newblock {\sl Complex angular momenta and the groups $SU(1,1)$ and $SU(2)$\/}, \newblock {\em Annals of Physics} {\bf 39}, 1--42 (1966); \\ Wayne J.~Holman and Lawrence C.~Biedenharn, \newblock {\sl A general study of the Wigner coefficients of $SU(1,1)$ \/}, {\em Annals of Physics} {\bf 47}, 205--231 (1968).

\bibitem{Wang1970} Kuo-Hsiang Wang, \newblock {\sl Clebsch-Gordan series and the Clebsch-Gordan coefficients of $O(2,1)$ and $SU(1,1)$\/}, \newblock {\em J. Math. Phys.} {\bf 11}(7), 2077--95 (1970).

\bibitem{Wybourne1974} B.~G.~Wybourne, \newblock {\sl Classical groups for physicists} (Wiley, New York, 1974).

\bibitem{Gerry2004} Christopher C.~Gerry, \newblock {\sl On the Clebsch-Gordan problem for $SU(1,1)$: coupling non-standard representations \/}, \newblock {\em J. Math. Phys.} {\bf 45}(3), 1180--90 (2004).

\bibitem{Gelfand1963} I.~Gel'fand, R.~Minlos and Y.~Shapiro, \newblock {\sl Representations of the rotation and Lorentz groups\/}
(Academic Press, New York, 1963).

\bibitem{LindbladNagel1970} G.~Lindblad and B.~Nagel, \newblock {\sl Continuous bases for unitary irreducible representations of $SU(1,1)$\/}, \newblock {\em Annales de l'Inst. Henri Poincar\'{e}} {\bf 13}, 27--56 (1970).

\bibitem{Castro2008} Carlos~Castro,
\newblock {\sl Born's reciprocal general relativity theory
and complex nonabelian gravity as gauge theory of
the quaplectic group : a novel path to quantum gravity\/},
 \newblock {\em Int. J. Modern
Phys.} {\bf A 23}(10), 1487--1506 (2008) .



\end{thebibliography}
\end{document}